\documentclass[fleqn,usenatbib]{mnras}

\usepackage{newtxtext,newtxmath}

\usepackage[T1]{fontenc}

\DeclareRobustCommand{\VAN}[3]{#2}
\let\VANthebibliography\thebibliography
\def\thebibliography{\DeclareRobustCommand{\VAN}[3]{##3}\VANthebibliography}

\usepackage[normalem]{ulem}
\usepackage{graphicx}	
\usepackage{amsmath}	
\usepackage{float}
\usepackage{natbib}
\usepackage{enumitem}
\hypersetup{breaklinks=true}

\title[Heterogeneous dust polarization model]{Modelling heterogeneous dust particles. An application to cometary polarization}
\author[P. Halder \& S. Ganesh]{
Prithish Halder$^{1}$\thanks{E-mail: prithishh3@gmail.com; prithish@prl.res.in}
and Shashikiran Ganesh,$^{1}$\thanks{E-mail: shashi@prl.res.in}
\\
$^{1}$Astronomy \& Astrophysics Division, Physical Research Laboratory, Ahmedabad, India\\
}

\date{
}

\pubyear{2020}

\begin{document}
\label{firstpage}
\pagerange{\pageref{firstpage}--\pageref{lastpage}}
\maketitle

\begin{abstract}
In this work, we introduce a comet dust model that incorporates multiple dust morphologies along with inhomogeneous mixture of silicate minerals and carbonaceous materials under power-law size distribution, to replicate the standard polarization-phase curve observed in several comets in the narrow-band continuum. Following the results from Rosetta/MIDAS and COSIMA, we create high porosity Hierarchical Aggregates (HA) and low porosity (< 10\%) Solids in the form of agglomerated debris. We also introduce a moderate porosity structure with solids in the core, surrounded by fluffy aggregates called Fluffy Solids (FS). We study the mixing combinations, (HA and Solids), (HA and FS) and (HA, FS and Solids) for a range of power-law index n=2.0 to 3.0 for different sets of mixing percentage of silicate minerals and carbonaceous materials. 
Polarimetry of the short period comets 1P/Halley and 67P/Churyumov-Gerasimenko match best with the polarisation resulting from the combination of HA and Solids
while the combinations (HA and FS) and (HA, FS and Solids) provide the best fit results for the long period comets C/1995 O1 (Hale-Bopp) and C/1996 B2 (Hyakutake). The best fit model results also recreate the observed wavelength dependence of polarization. Our dust model agree with the idea that the long period comets may have high percentage of loose particles (HA and FS) compared to those in the case of short period comets as the short period comets experience more frequent and/or higher magnitude of weathering.        

\end{abstract}

\begin{keywords}
polarization -- radiative transfer -- software:simulations -- comets: general -- comets: 1P/Halley -- comets: 67P/Churyumov-Gerasimenko -- comets: C/1995 O1 (Hale-Bopp) -- comets: C/1996 B2 (Hyakutake)
\end{keywords}



\section{Introduction}

Comet dust particles are some of the pristine materials present in our solar system. They hold the primordial signatures of the chemical and physical processes that led to their formation in the solar proto-planetary disk. The study of electromagnetic scattering of sunlight by cometary dust particles provides crucial information about the physical properties 
of the dust. The scattering of sunlight by cometary dust particles 
ensures that the light from the coma of a comet is polarised. Fluorescence emission by gaseous molecules in the cometary coma depolarises this to some extent.  The degree of linear polarization is one of the significant light scattering parameters that has been extensively studied, by several astronomers in the past, through observations of comets{\citep{kiselev2005,joshi2010,debroy2015,hadamcik2016}}.

The observed polarization in comets depend on the size, shape, and composition of the dust particles, the wavelength of observation and the Sun-Comet-Earth phase angle at the time of observation.  These observations are used to generate the polarization-phase curve.  The polarization-phase curves of different comets show some similar characteristics, $viz.$ a bell shaped curve with maximum positive polarization in the phase angle range  90$^{\circ}$ to 110$^{\circ}$ having polarization percentage within 10\% to 30\% {\citep{Levasseur-Regour1996}}, a distinct negative polarization branch in the back-scattering region having inversion angle around 20$^{\circ}$ with minimum polarization around -2\% {\citep{dollfus1988,eaton1992}}. These characteristics are observed in several comets irrespective of their size or activity. 
However, the maxima of the positive polarization and the minima of the negative polarization 
varies for different comets {\citep{kolokolova2004}}.

Recent paper by {\citet{rosenbush2017}} is a good example of the combination of observation and theoretical modelling which explains the observed spatial variation of polarization, colour and brightness of comet 67P/Churyumov-Gerasimenko with silicate and organic grains having a power law distribution spanning a large range of sizes.

In the light scattering simulations of comet dust particles, the dust is often modelled as aggregates of monodisperse and homogeneous spheres {\citep{mukai1991,joshi1997,petrova2000,kimura2006,shen2009}}. \citet{lumme2011} efficiently modelled the observed polarimetric response from comets using  polydisperse aggregates (polydispersity in monomer size as well as aggregate size). 
The findings from Rosetta/MIDAS {\citep{mannel2019}} suggests that the smallest sub-unit/grain size of an aggregate is approximately 0.1$\mu$m and has a log normal size distribution and standard deviation of $\pm$ 0.020$\mu$m to $\pm$ 0.035$\mu$m. This size standard deviation indicates the presence of a substantial amount of polydispersity in the monomers/sub-units at sub-micron level of an aggregate. 
On the other hand, both the Rosetta and Stardust missions have indicated the presence of a solid group of particles with low porosity (porosity $<$ 10\%) \citep{guttler2019}. Modelling cometary polarization using such solid particles in the form of agglomerated debris by {\citep{zubko2011, zubko2012, zubko2014, zubko2016, zubko2020}} has confirmed that the solid group of particles play a crucial role in the overall polarimetric response from comets. 

With increasing particle size, the surface area as well as the effective volume of the inclusions increases and as it reaches to a size comparable to the wavelength of incident radiation, the inhomogeneity due to the presence of inclusions increases. The effect of inhomogeneity is studied by {\citet{markkanen2015}} \& \citet{videen2015} where they used inhomogeneous irregular-particle model to study the light scattering properties.

The Rosetta findings also indicate the presence of both porous aggregates and solid (low porosity) group of particles. Moreover the aggregates are mainly found to be hierarchical in nature, made up of sub-structures or sub-aggregates which are further made up of polydisperse sub-units or monomers. \citet{zubko2005} studied the intensity and the degree of linear polarization of light scattered by hierarchical debris particles by considering different hierarchical levels from densly packed debris to porous debris over a range of sizes. On the other hand the polarimetric and photometric response from hierarchical structures of monodisperse aggregates are studied by \citet{kolokolova2018} and they have stated that comet dust models using aggregates must consider hierarchical structures. Hence, it is required to explain the average properties of comets depending on the observed and measured set of constraints that incorporates inhomogeneity, grain polydispersity, hierarchical structures and solid group of particles. 

Finally, a comet dust model must reflect the presence of morphologically different particles based on porosity as studied by \citet{das2011} \& \citet{dlugach2018}.

Thus, in this work, we propose a generalised comet dust model that uses different particle morphologies and inhomogeneous mixture of silicates \& carbonaceous materials 
for varying power-law size distribution index to explain the observed polarimetric phase function as well as the observed wavelength dependence of polarization for different comets.  
We consider a heterogeneous system of aggregated particles having inhomogeneity in refractive indices as well as polydispersity of sub-unit/grain size (based on the measured morphology). We also consider the low porous solid group of particles in the form of agglomerated debris and introduce a new morphological class as a mixture of hierarchical aggregates (HA) and solids, termed as fluffy solids (FS), in order to explain the observed optical polarization. The Hieararchical aggregates considered in our model  differs from that of \citet{zubko2005} in terms of the structural morphology of the base particles considered. We consider fluffy BCCA aggregates in place of the low porous agglomerated debris used by them.

The layout of the different sections of the paper is as follows.  In Section 2.0 we discuss the modelling methodology.  We explain the different model dust morphologies in section 2.1, composition in 2.2 and numerical simulations in 2.3. In section 3.0 we provide the light scattering results. Finally, in section 4.0, we compare our model results with the comets 67P/C-G in 4.1, 1P/Halley in 4.2, C/1995 O1 (Hale-Bopp) in 4.3 \& C/1996 B2 (Hyakutake) in 4.4 followed by discussion in section 5 \& conclusions in section 6.

\section{Modelling methodology}

In this section we describe the different particle morphologies considered in modelling the cometary dust. 
Short period comets experience higher magnitude of weathering of loose particles compared to that in the case of long period comets. Thus, we expect that the structural morphology of the dust particles may vary from comet to comet on the basis of orbital period or dynamical age.
So, we introduce three different kinds of structural morphologies on the basis of porosity, $viz.$ hierarchical aggregates (HA) of high porosity, solids with low porosity and fluffy solids (FS) with moderate porosity as they are mixtures of solids and aggregates.  Initially, we study the fractal aggregates having hierarchical structure in section 2.1.1 \& 2.1.2, then in section 2.1.3 we discuss the low porosity solid particles as agglomerated debris and finally we introduce the fluffy solids in the section 2.1.4. The HA are built up of polydisperse sub-units/monomers using Ballistic Cluster Cluster Agglomeration method. The low porous solids are made up of agglomerated debris particles, while the fluffy solids are made up of mixture of solids and porous hierarchical aggregates. The three different morphologies indicate different levels of porosities ($i.e$ high, moderate and low porosity) which correspond to different bulk densities in the range (0.3-3.0) g/cm$^3$ \citep{hortz2006} . 
Compositionally, we consider different mixing percentages of silicate minerals and carbonaceous materials in each of the different dust morphologies.
After building up the model dust material, we carry out numerical simulations of the scattering phenomena.

We have used the Vikram-100 High Performance Computing (HPC) facility at PRL, Ahmedabad, to perform all the numerical simulations used in this study. The speed of each simulation is highly sensitive to the structural morphology, size and composition of the model structure. The largest structure considered in this study used 500 out of the 2328 CPU cores. 

\subsection{Model dust components}
\subsubsection{Fractal aggregates}
\label{sec:fractal} 
Interplanetary dust particles (IDPs) have been collected from Antarctic ice as well as in the Earth's stratosphere{\citep{brownlee1985,noguchi2015}}.  They show that the IDPs are constituted by porous anhydrous chondrites.  They are agglomerates of smaller grains or sub-units, also called primary particles(PP) {\citep{lawler1992, bradley2003}}.  The collected IDPs are carbonaceous organic-rich materials which are believed to originate from short period comets. The studies by \citet{Kimura2003a} indicate a high resemblance of the chemical composition of IDPs with those found from the in-situ measurements of the comet 1P/Halley.  
\begin{figure}
	\centering
	\includegraphics[width=\columnwidth]{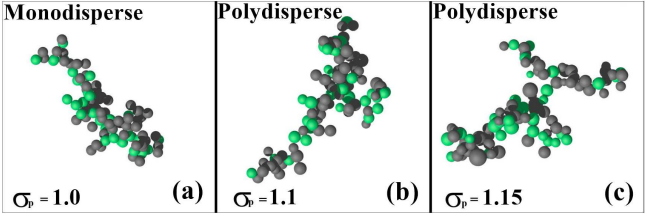}
    \caption{Inhomogeneous aggregate having 70\% carbonaceous material {\textbf{(grey spheres)}} and 30\% silicate minerals {\textbf{(green spheres)}}, $N$ = 100, $D_f$ = 1.8 and $x_p$ = 1.4 for varying polydispersity (a) $\sigma_p$ = 1.0, (b) $\sigma_p$ = 1.1 \& (c) $\sigma_p$ = 1.15} \label{fig:polydispersity}
\end{figure}
While a random arrangement of primary particles is considered an aggregate, agglomerates are loosely bound clumps of primary particles, aggregates of primary particles or a mixture of both.  A dust aggregate is a result of grain-grain collisions of pristine submicron condensates. 
The structural morphology of such dust aggregates is defined by fractal geometry due to the low relative velocity of grain-grain collisions in the early phase of dust growth.   

The structure of such an aggregate of \textit{N} monodisperse grains having grain radius \textit{r} is defined by the following equation {\citep{mandelbrot1983}},
\begin{equation}
     N = k_{f}\left(\frac{R_{g}}{r}\right)^{D_{f}}
 	\label{eq:quadratic}
\end{equation}

\noindent where $R_g$ is the radius of gyration, $D_f$ is the fractal dimension of the aggregate and $k_f$ is the fractal prefactor.

The radius of gyration, $R_g$, is defined by the following equation
\begin{equation}
    R_g = \left[\frac{1}{2N^{2}}\sum_{i}^{N}\sum_{j}^{N}\left(\vec{r_i} - \vec{r_j}\right)^{2}\right]^{1/2}
\end{equation}

\noindent where $\vec{r_i}$ and $\vec{r_j}$ are the position vectors of the $i^{th}$ and $j^{th}$ grains {\citep{filippov2000}} 

The apparent radius, or the characteristic radius, of the aggregate is defined by
\begin{equation}
    R_c = \sqrt{\frac{5}{3}}R_g
\end{equation}

In the case of polydisperse grains having PP radius $r_p$, average PP mass $\overline{m_p}$ and aggregate mass $m_a$, the structural morphology of the aggregate is described by {\cite{eggersdorfer}} as 
\begin{equation}
    \frac{m_a}{\overline{m_p}}= k_{f}\left(\frac{R_{g}}{r_p}\right)^{D_{f}}
\end{equation}

\noindent where $m_a/\overline{m_p} = N$ is the number of PP in the aggregate.

In this study we considered log-normal PP size standard deviation $\sigma_p$. Figure \ref{fig:polydispersity} depicts the morphology of inhomogeneous aggregate with changing polydispersity $\sigma_p$ = 1.0, 1.1 \& 1.15.

\subsubsection{Hierarchical aggregate structure}

The studies from Rosetta/COSIMA and Rosetta/MIDAS instruments  {\citep{guttler2019,mannel2019}} suggest a hierarchical, porous structure of dust aggregates having sub-aggregates or sub-structures made up of fluffy sub-units. They also suggested that the fluffy sub-units have mean primary particle radius of 0.1$\mu$m  and size standard deviation of $\pm$ 0.02$\mu$m  to $\pm$ 0.035$\mu$m. The smallest identified sub-units have sizes between 0.05$\mu$m and 0.2$\mu$m. We build such hierarchical polydisperse inhomogeneous aggregate as cometary dust candidates (scatterer) to be used in the light scattering simulations.  We consider PP (or monomer or sub-unit) size parameter, $x_p$, given by:

  \begin{equation}
    x_p = \frac{2\pi}{\lambda}r_p = 1.4
    \begin{cases}
      r_p = 0.152\micron, & \text{for}\ \lambda = 0.6840~\micron \\
      r_p = 0.109\micron, & \text{for}\ \lambda = 0.4845~\micron \\
      r_p = 0.081\micron, & \text{for}\ \lambda = 0.3650~\micron
      
    \end{cases}
  \end{equation}
\begin{figure*}
	\includegraphics[width=\textwidth]{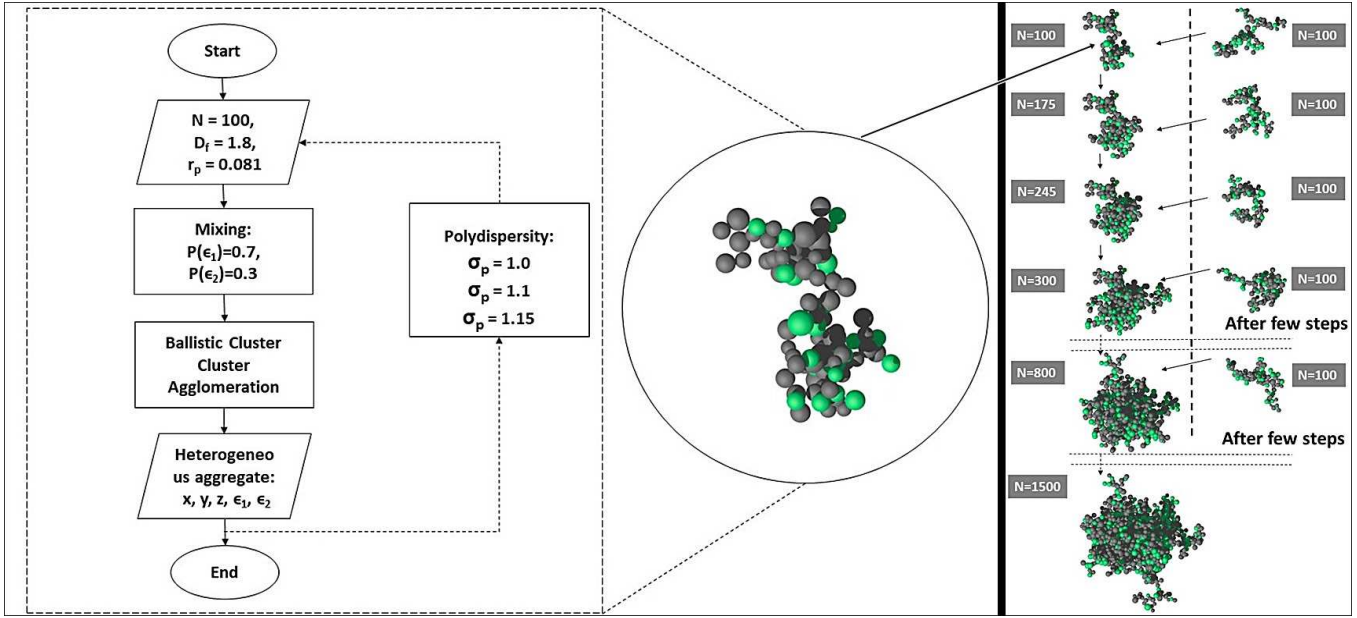}
    \caption{Flowchart of the construction of the aggregate from the primary particle.}
    \label{fig:flowchart}
\end{figure*}
In order to create polydisperse inhomogeneous aggregate we used the package known as FLAGE\footnote{FLAGE is available for download from \url{scattering.eu}}{\citep{Skorupski2014}}. For a particular set of input parameters defining the number of monomers/sub units ($N$), fractal dimension ($D_f$), fractal prefactor ($k_f$), radius of PP ($r_p$), log-normal PP size standard deviation ($\sigma_p$) and volume concentration of different materials, one can create a polydisperse inhomogeneous aggregate.  
The initial aggregate that we considered has $N$ = 100 sub-units and $R_g$ = 0.9\micron. This aggregate is then ballistically collided with another morphologically similar aggregate having 100 number of sub-units to form a final aggregate. The resultant aggregate is then collided with another similar aggregate having 100 sub-units. This procedure is repeated for each resultant aggregate till we reach a desirable size. Each of the resultant aggregate is used as a candidate for light scattering simulations. Figure \ref{fig:flowchart} depicts the flow diagram of the entire procedure.

\subsubsection{Solids}

Both Rosetta/MIDAS as well as Rosetta/COSIMA \citep{guttler2019} reported substantial amount of negligibly porous solid particles. In this model we introduce solid particles in the form of agglomerated debris particles which has 23.6\% of particle volume fraction. We developed a simple code to construct such particles, following the algorithm  specified by {\citep{zubko2006}} to break a spherical target of $N$ dipoles by randomly selecting dielectric points for vacuum and material.

\begin{figure*}
	\includegraphics[width=\textwidth]{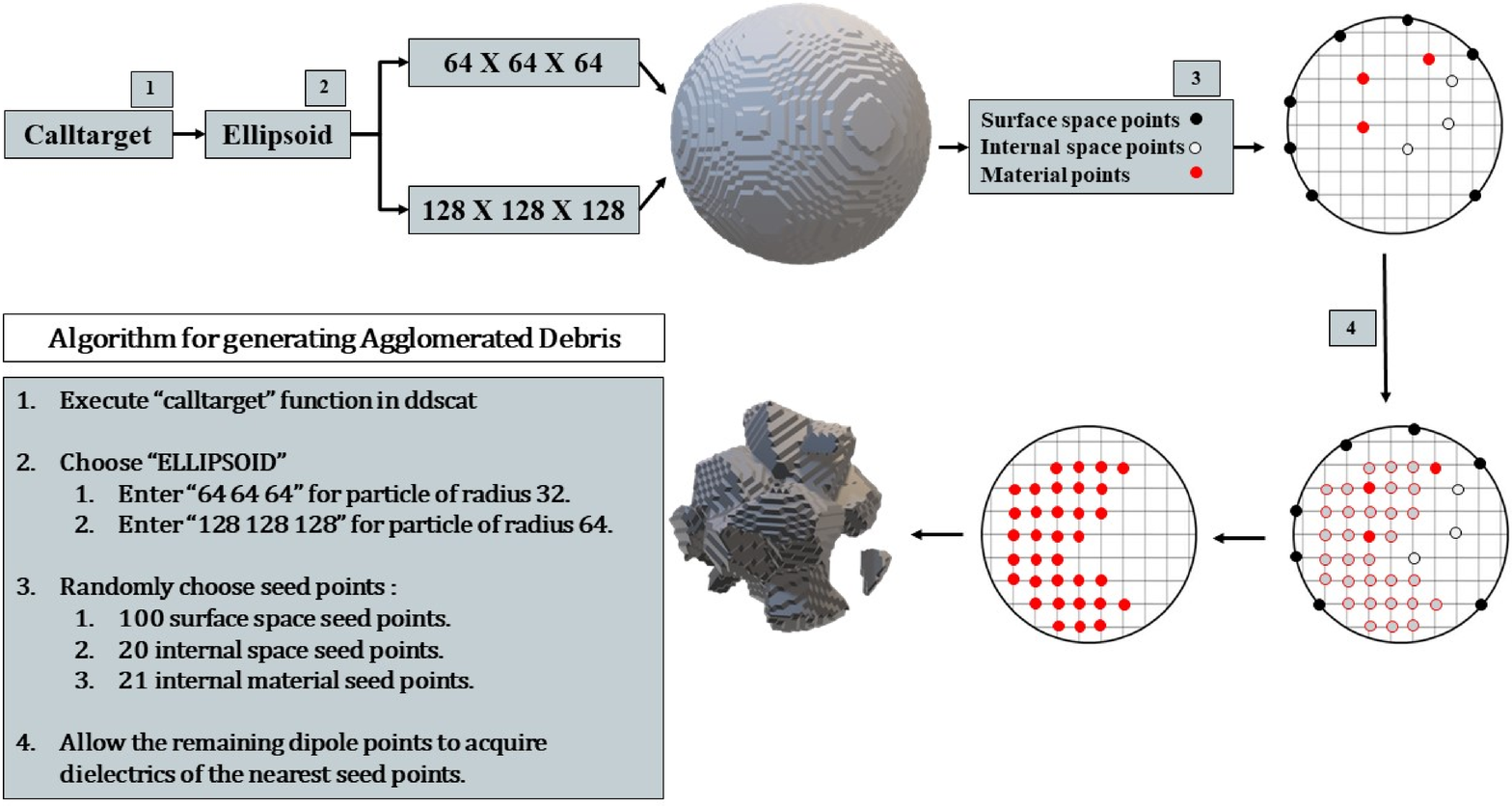}
    \caption{Algorithm for generating aglomeratted debris (Solid) particles as obtained from \citet{zubko2006}}
    \label{fig:AD_algorithm}
\end{figure*}

\begin{figure}
	\includegraphics[width=\columnwidth]{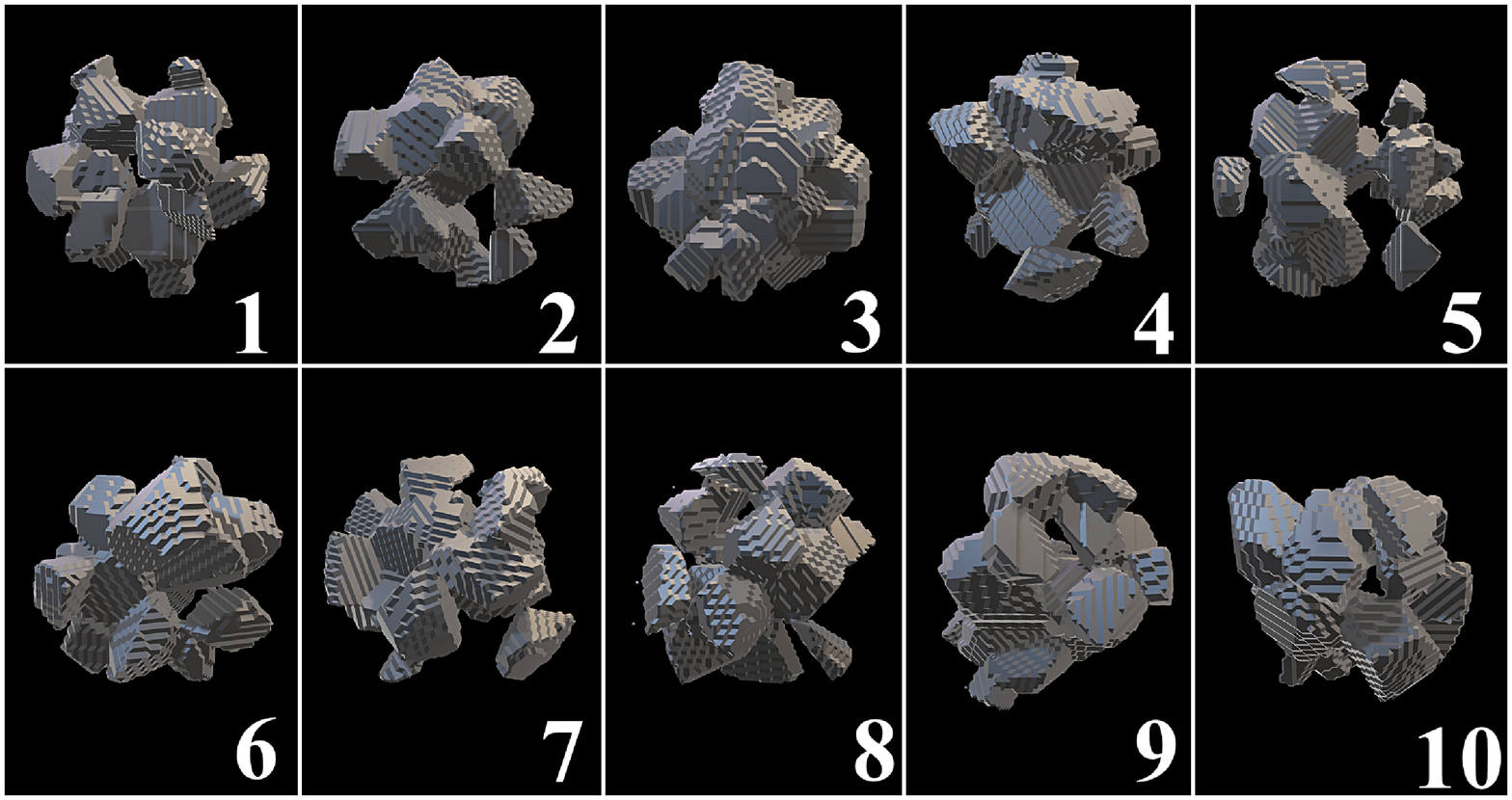}
    \caption{Ten realisations of the agglomerated debris particles.}
    \label{fig:solids}
\end{figure}

To generate agglomerated debris we create spherical target of 137,376 dipoles for size parameters (x) < 16 and target of 1,099,136 dipoles for x = 16 to 32 \citep[numbers taken from][]{zubko2006,zubko2020} by using "CALLTARGET" function in "DDSCAT". We choose 100 random points from the surface of the sphere as surface space seed points. In the internal volume of the sphere we randomly choose 20 and 21 seed points for the space/vacuum and material respectively. After selecting the seed points, each dipole points other than the seed points are allowed to acquire the dielectrics of the nearest seed points. Figure \ref{fig:AD_algorithm} depicts the algorithm for generating irregular agglomerated debris particles. Figure \ref{fig:solids} exhibits the ten different realisations of the agglomerated debris particles that are being used as a representation of the solid particles. In this study we use \emph{Solids Realisation 1} (depicted in Figure \ref{fig:solids}) as our model structure for the calculation of light scattering properties of Solids. It must be noted that agglomerated debris particles are actually collection of several debris located at few distances from each other. Hence it will be worth studying the effect of distance between these debris particles on the light scattering response. However, this is an open problem for future studies related to agglomerated debris particles.

\subsubsection{Fluffy Solids}

\begin{figure}
	\includegraphics[width=8cm]{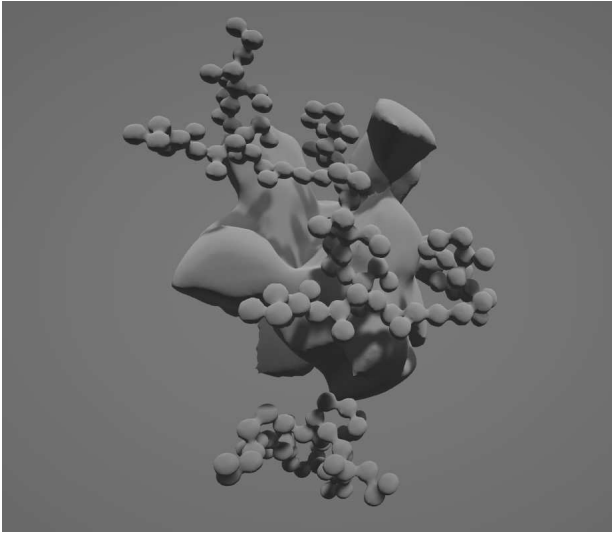}
    \caption{Example structure of FLUFFY SOLIDS (FS)}
    \label{fig:fluffy_solds}
\end{figure}

Apart from HA and Solids we introduce a mixed morphology of solid and fluffy aggregates called Fluffy Solids (FS) which are significantly loose particles.  We expect to see such particles in the case of dynamically new comets since these particles were observed in IDPs. Images of the IDPs are recorded in the NASA Cosmic Dust Catalog\footnote{L2021A7 from the catalog available at the url \url{https://curator.jsc.nasa.gov/dust/cdcat15/cdcat15_combined.pdf}}. We separately created agglomerated debris and fluffy aggregates and mixed them using Blender3D\footnote{Blender3D is available for free download under the General Public License (GPL) at \url{https://www.blender.org/}}, keeping the solid in the core surrounded by fluffy aggregates as filaments as shown in Figure \ref{fig:fluffy_solds}.
\subsection{Dust model}
We consider polydisperse and inhomogeneous particles having hierarchical aggregate structure with different morphologies, as described in the previous sub-section, to be our model dust particles. The polydispersity of the aggregate is controlled by the log-normal PP size standard deviation $\sigma_p$. As we increase $\sigma_p$ from 1.0 to 1.15 the polydispersity increases from 0$\mu$m to $\pm$ 0.04 $\mu$m. 
We consider an inhomogeneous mixture of carbonaceous materials and silicate minerals, which are believed to be the most abundant materials in comets. The refractive index of silicate minerals {\citep{scott1996}} and carbonaceous materials {\citep{jenniskens1993,li1997}} are listed below:

  \begin{equation}
    Silicate
    \begin{cases}
      1.677+0.004\emph{i}, & \text{for}\ \lambda = 0.6840~\micron \\
      1.684+0.003\emph{i}, & \text{for}\ \lambda = 0.4845~\micron \\
      1.700+0.002\emph{i}, & \text{for}\ \lambda = 0.3650~\micron
    \end{cases}
  \end{equation}

  \begin{equation}
    Carbonaceous
    \begin{cases}
      1.99+0.222\emph{i}, & \text{for}\ \lambda = 0.6840~\micron\\
      1.97+0.231\emph{i}, & \text{for}\ \lambda = 0.4845~\micron \\
      1.97+0.236\emph{i}, & \text{for}\ \lambda = 0.3650~\micron
    \end{cases}
  \end{equation}

The material inhomogeneity is introduced in the aggregate by randomly selecting the refractive index parameter with the probability $P(\epsilon_i)$ for each sub-unit, where $\epsilon_i$ is the refractive index of the $i$th composition. The probability of carbonaceous and silicate materials used in the inhomogeneous aggregates are $P(\epsilon_1)$ = 0.70 (70\% of carbonaceous materials $\epsilon_1$) and $P(\epsilon_2)$ = 0.30 (30\% of silicate minerals $\epsilon_2$) respectively.

The fractal dimension associated with the modelled dust aggregate are considered to be $D_f$ = 1.8. This value is consistent with the recent observations that suggests $D_f$ < 2 {\citep{mannel2019}}. 

In case of solid agglomerated debris particles, we consider homogeneous composition, separately for silicate minerals and carbonaceous particles.

We calculate the light scattering properties of hierarchical aggregates using the multi-sphere T-matrix (MSTM-V3.0) technique and for the solid agglomerated debris particles we use Discrete Dipole Approximation (DDSCAT-7.3) as described in the following subsection.
The size range of HA used in this model is 1.5$\micron$ - 6.8$\micron$ and in the case of Solids the size range is 0.11$\micron$ - 3.41$\micron$.

\subsection{Numerical simulations}
The theoretical model of light scattering phenomenon in the coma of a comet is primarily based on the premise of a low volume concentration  of comet dust particles. We consider the coma to be optically thin and ignore multiple scattering effects. The scattering  phenomenon for a mirror symmetric and macroscopically isotropic particulate medium is defined by the phase matrix which represents far field transformation of the Stokes parameters of the incident light ($I_i$, $Q_i$, $U_i$, $V_i$) to that of the scattered light ($I_s$, $Q_s$, $U_s$, $V_s$).   This phase matrix is given by \citep{bohren2008}:
\begin{equation}
\left( \begin{array}{cccc}
I_s  \\
Q_s  \\
U_s  \\
V_s  \end{array} \right)= \frac{1}{k^{2}d^{2}} \left( \begin{array}{cccc}
    S_{11} & S_{12} & 0 & 0 \\
    S_{12} & S_{22} & 0 & 0 \\
    0 & 0 & S_{33} & S_{34} \\
    0 & 0 & -S_{34} & S_{44} \end{array} \right)\left( \begin{array}{cccc}
I_i  \\
Q_i  \\
U_i  \\
V_i  \end{array} \right)
\end{equation}
where $k$ is the wave-number and $d$ is the distance between the scatterer and the observer and $S_{ij}$ represents the orientationally symmetric scattering matrix elements. Due to mirror and rotational symmetry each of the scattering matrix elements depend on the scattering angle  and are independent of the azimuthal angle. The angle $\alpha$ between Sun-Comet-Earth is called the \emph{Phase angle}. Angle $\theta$ = 180 - $\alpha$ is called the \emph{Scattering angle}, $\theta$ = [0$^{\circ}$,180$^{\circ}$]. 

In this work we study the following light scattering parameters defined by the scattering matrix elements taken from {\cite{bohren2008}}:

\begin{enumerate}
    \item \emph{Phase function}: $S_{11}$  
    
    \item \emph{Degree of linear polarization}: \(DP = -S_{12}/S_{11}\).
    
    \item \emph{Depolarization ratio}: \(D_{1} = 1 - \frac{S_{22}}{S_{11}}\)
    
    \item \emph{Anisotropy parameter}: \(D_{2} = \frac{S_{33}}{S_{11}} - \frac{S_{44}}{S_{11}}\)
\end{enumerate}
   
The last two parameters indicate the anisotropies present in the scatterer. The anisotropy condition for a non spherical scatterer is, \(S_{11} \neq S_{22}\) and \(S_{33} \neq S_{44}\).

The above parameters are obtained from the numerical simulations of electromagnetic scattering of sunlight from hierarchical aggregate particles.  For this we use the parallel version of the Multi-sphere T-matrix (MSTM) code\footnote{MSTM code is available for download from \url{http://eng.auburn.edu/users/dmckwski/scatcodes/}} developed by {\cite{mackowski2011}}. 

We use the Discrete Dipole Approximation(DDA) scattering codes\footnote{Discrete Dipole Approximation (DDA) DDSCAT version 7.3.3 \url{http://ddscat.wikidot.com/}} {\citep{draine1994}} in parallel mode, for the numerical simulations of light scattering by solid agglomerated debris particles.

Further, the results are averaged using the power-law size distribution $r^{-n}$ where $n$ ranges between 2.0 to 3.0. 

\begin{figure*}
	\centering
	\includegraphics[width=15cm]{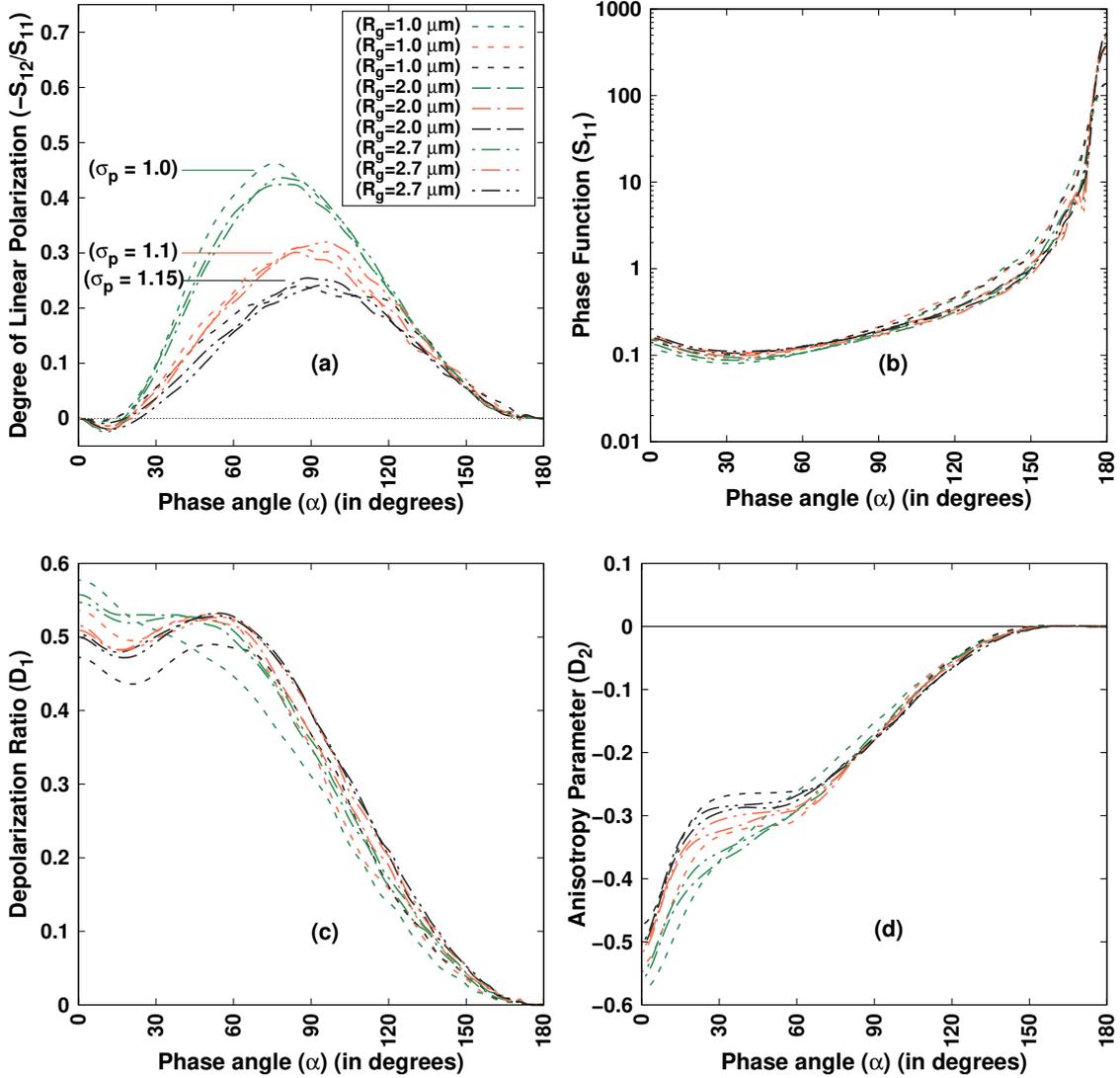}
    \caption{Effect of variation of $\sigma_p$ (with values 1.0, 1.1 and 1.15) on (a) Degree of linear polarization ($-S_{12}/S_{11}$), (b) phase function ($S_{11}$), (c) depolarization ratio ($D_1$) and anisotropy parameter ($D_2$) as function of phase angle
    }
    \label{fig:sigma}
\end{figure*}

\section{Light scattering results}
In this section we present the results of the light scattering simulations for Hierarchical aggregates (HA), Solids \& Fluffy Solids (FS).

\subsection{Hierarchical Aggregates (HA)}

\subsubsection{Effect of polydispersity $\sigma_p$}
Polydispersity refers to the geometric standard deviation of the size of the PP ($\sigma_p$). In this study we use three different values of $\sigma_p$ i.e. 1.0, 1.1 and 1.15. The corresponding normal standard deviation, $\sigma$, are 0$\micron$, $\pm$0.03$\micron$ and $\pm$0.04$\micron$.

  \begin{equation}
    \sigma_p
    \begin{cases}
      = 1.0, & \text{for Monodisperse,}\  \sigma = 0 \micron  \\
      = 1.1, & \text{for Polydisperse,}\  \sigma = \pm 0.03 \micron \\
      = 1.15, & \text{for Polydisperse,}\  \sigma = \pm 0.04 \micron
    \end{cases}
  \end{equation}
  

\begin{figure*}
	\centering
	\includegraphics[width=15cm]{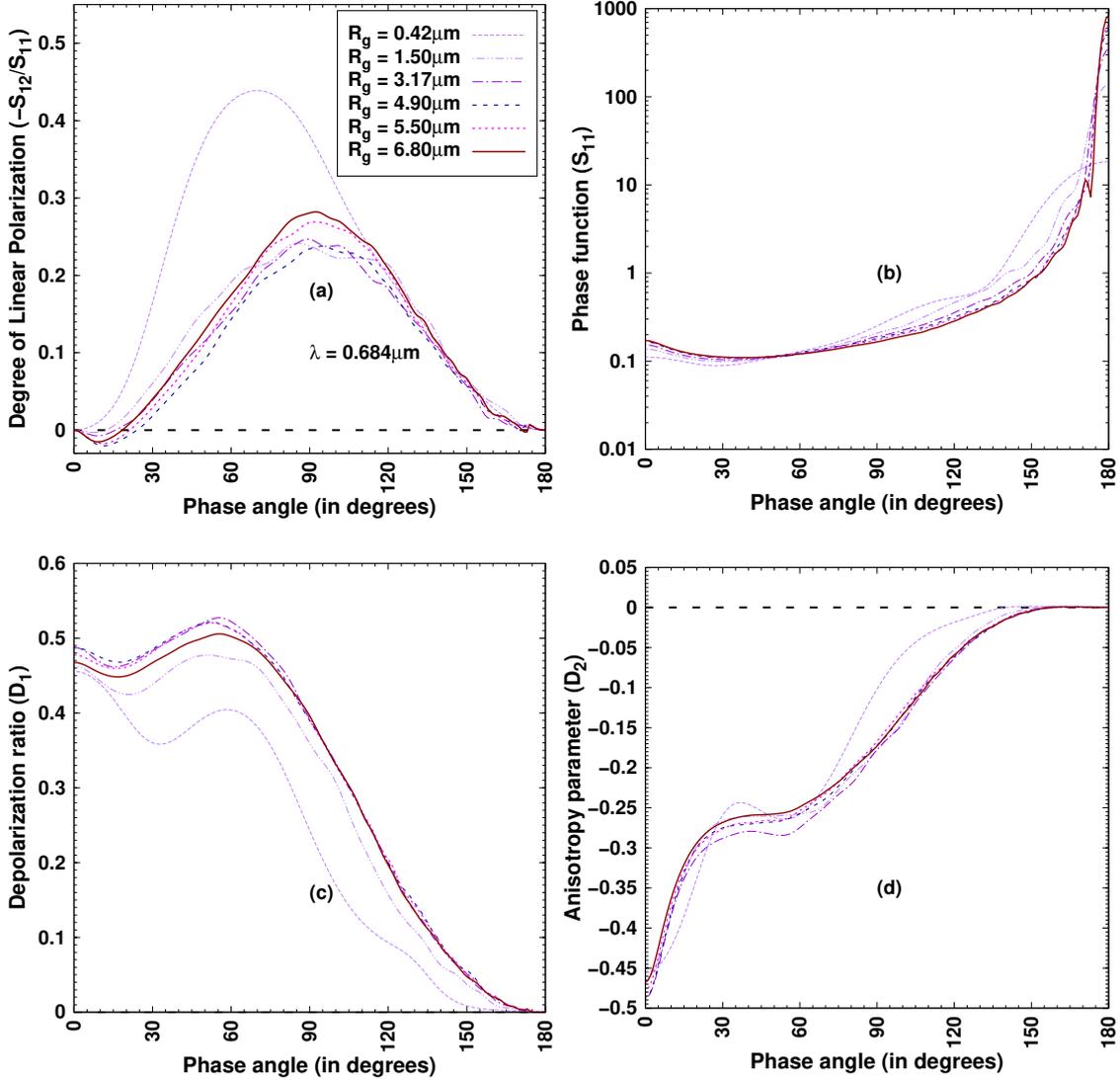}
    \caption{Effect of increasing aggregate size $R_g$ on (a) Degree of linear polarization ($-S_{12}/S_{11}$), (b) phase function ($S_{11}$), (c) depolarization ratio ($D_1$) and anisotropy parameter ($D_2$) as a function of phase angle.} 
    \label{fig:sizeparameter}
\end{figure*}
Figure \ref{fig:sigma} shows the effect of changing polydispersity on the light scattering parameters, 
for three aggregates having different values of $R_g$ (1.0$\micron$, 2.0$\micron$ and 2.7$\micron$) at wavelength $\lambda$=0.6840$\micron$.
The maximum of the degree of linear polarization, $P_{max}$, shows an abrupt decrease of $\approx$ 20\% with increasing polydispersity $\sigma_p$. Also the slope of the polarization curve decreases with increasing $\sigma_p$.
The phase function ($S{11}$) shows negligibly small variation with increasing $\sigma_p$, whereas,  the depolarization ratio ($D_1$) and anisotropy parameter ($D_2$) show a significant variation with increasing $\sigma_p$ in the backscattering region. However, $D_2$ develops a peak  with increasing $\sigma_p$ around the phase angle 30$^{\circ}$. This indicates a sudden increase in the anisotropy of the scattering target with increasing polydispersity. 

The abrupt variation in the degree of linear polarization and the anisotropies in the backscattering region can be attributed to the cumulative effect of changing $\sigma_p$. As all other physical properties of the aggregate are fixed, the changing $\sigma_p$ increases multiple scattering gradually in the low phase angles as shown in Figure \ref{fig:sigma}(c) \& \ref{fig:sigma}(d).

\cite{halder2018} discussed the comparison between different light scattering parameters with changing porosity, aggregate size, monomer size and composition. The study indicates the presence of an induced relation between the degree of linear polarization \& the associated anisotropies (D$_{1}$ \& D$_{2}$). A small change in the real part of the refractive index exhibits variation in the degree of linear polarization which is accompanied by variation in the anisotropies, whereas the corresponding change in the phase function is comparatively less. However, a small change in porosity, size \& imaginary refractive index shows noticeable variation in the phase function.

\subsubsection{Effect of aggregate size $R_g$}
The size of an aggregate is a very crucial factor as all the light scattering parameters show significant variation with increasing aggregate size. In this section we study the effect of increasing aggregate size $R_g$ of the model hierarchical polydisperse inhomogeneous dust aggregates for a fixed polydispersity $\sigma_p$.
\begin{figure*}
	\includegraphics[width=\textwidth]{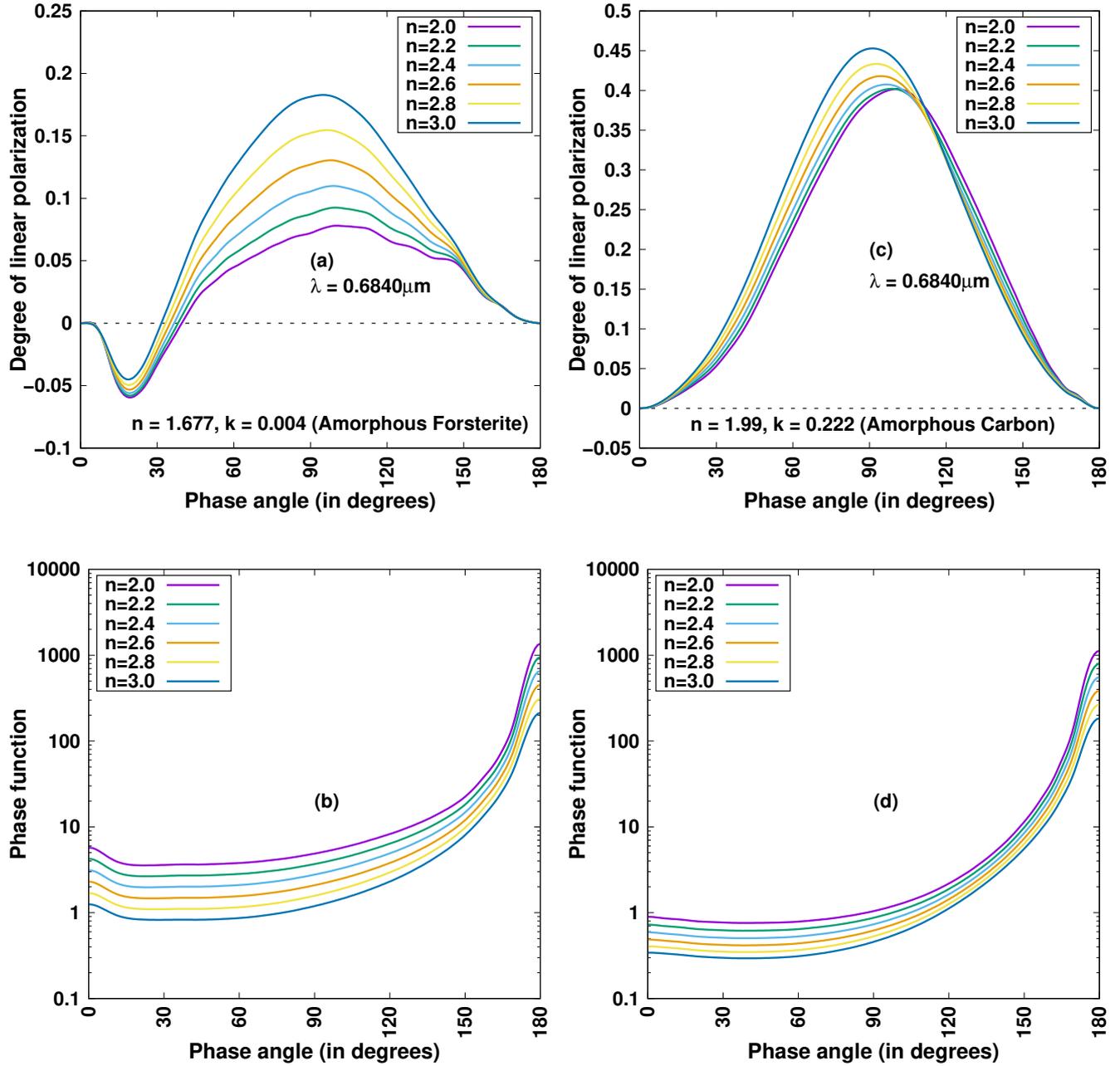}
    \caption{Variation of the degree of linear polarization and phase function with increasing power-law index for amorphous forsterite (a)\&(b) and Amorphous carbon (c)\&(d) respectively at $\lambda$ = 0.6840$\mu$m .}
    \label{fig:ad_pol_phase}
\end{figure*}
Figure \ref{fig:sizeparameter} shows ($a$) the variation of the degree of linear polarization (-$S_{12}/S_{11}$), ($b$) phase function ($S_{11}$), ($c$) depolarization ratio ($D_1$) and ($d$) anisotropy parameter ($D_2$ ) with increasing $R_g$ for $\sigma_p$ = 1.15 at the narrow band wavelength $\lambda$ = 0.6840 $\mu$m. With increasing size, there is a significant reduction in the value of polarization till $R_g$ $\approx$ 3.0$\mu$m to 5.0$\mu$m and beyond that, the value of polarization increases. A similar trend is observed in the laboratory measurements by {\citet{hadamcik2009}}, which shows a decreasing polarization trend with increasing particle size till 4.0$\mu$m and beyond that polarization increases. Though in our calculations the highest aggregate size is $R_g$ $\approx$ 7$\mu$m which is < 10$\mu$m, yet the sudden shift in polarization is distinct. The shift is very much prominent in the parameter $D_2$ where the peak around phase angle 30$^{\circ}$ attains highest value for $R_g$ $\approx$ 3.0$\mu$m to 5.0$\mu$m and reduces beyond that. This shift in polarization is also detected in the spatial variation of polarization in the comet 67P/Churyumov-Gerasimenko {\citep{rosenbush2017}}. We compare results from our model with the observations of 67P/Churyumov-Gerasimenko in section 4.2.
\subsection{Solids}
\begin{figure*}
	\includegraphics[width=\textwidth]{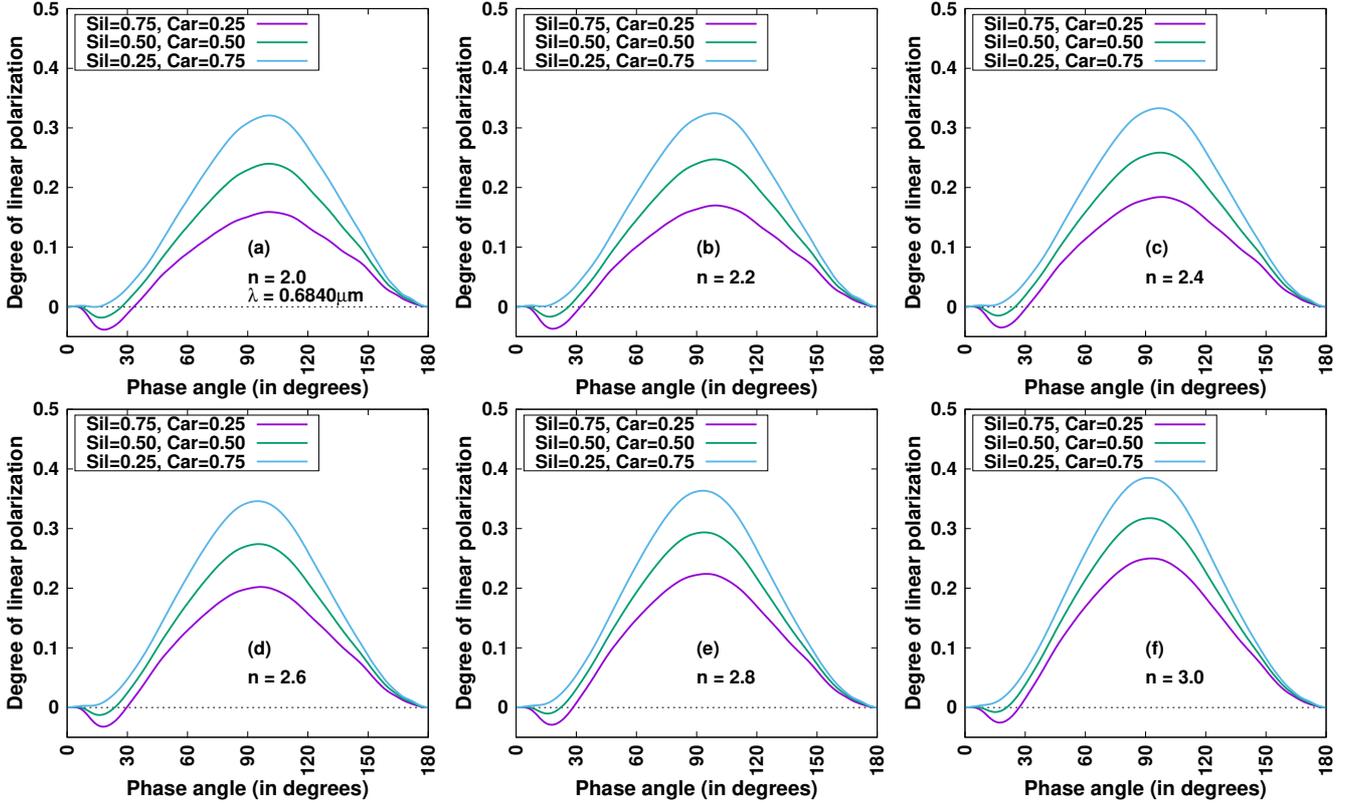}
    \caption{Variation of the degree of linear polarization for Solids (agglomerated debris) with different mixing percentage of silicate and carbon for increasing power-law index $n$ = 2.0 (a), 2.2 (b), 2.4 (c), 2.6 (d), 2.8 (e) \& 3.0 (f) respectively at $\lambda$ = 0.6840$\mu$m .}
    \label{fig:ad_mix_S_C}
\end{figure*}

In this section we present the results of our light scattering simulations for the solid group of particles (agglomerated debris particles with low porosity for silicate minerals and carbonaceous materials respectively.
Figure \ref{fig:ad_pol_phase} shows the variation in polarization and phase function with changing power-law size distribution index $n$ for silicate minerals and carbonaceous materials respectively at $\lambda$ = 0.6840 $\mu$m. 
The degree of linear polarization, Figures  \ref{fig:ad_pol_phase}{(a) \& (c)}, exhibits a broader variation with $n$, in case of silicate, whereas in case of carbonaceous materials the variation tends to be shallow. Thus, keeping in mind the sensitivity of polarimetric response towards the silicate materials, we mix the values obtained for silicates and carbonaceous materials for different mixing ratios as shown in Figure \ref{fig:ad_mix_S_C}. It is clear from the figure, that the degree of linear polarization increases with decreasing silicate percentage. 
Further, the variation with $n$ from Figure \ref{fig:ad_mix_S_C} reveals that the degree of linear polarization is higher for small size particles whereas it tends to be shallower for large size particles.
\subsection{Fluffy Solids (FS)}
\begin{figure*}
	\includegraphics[width=\textwidth]{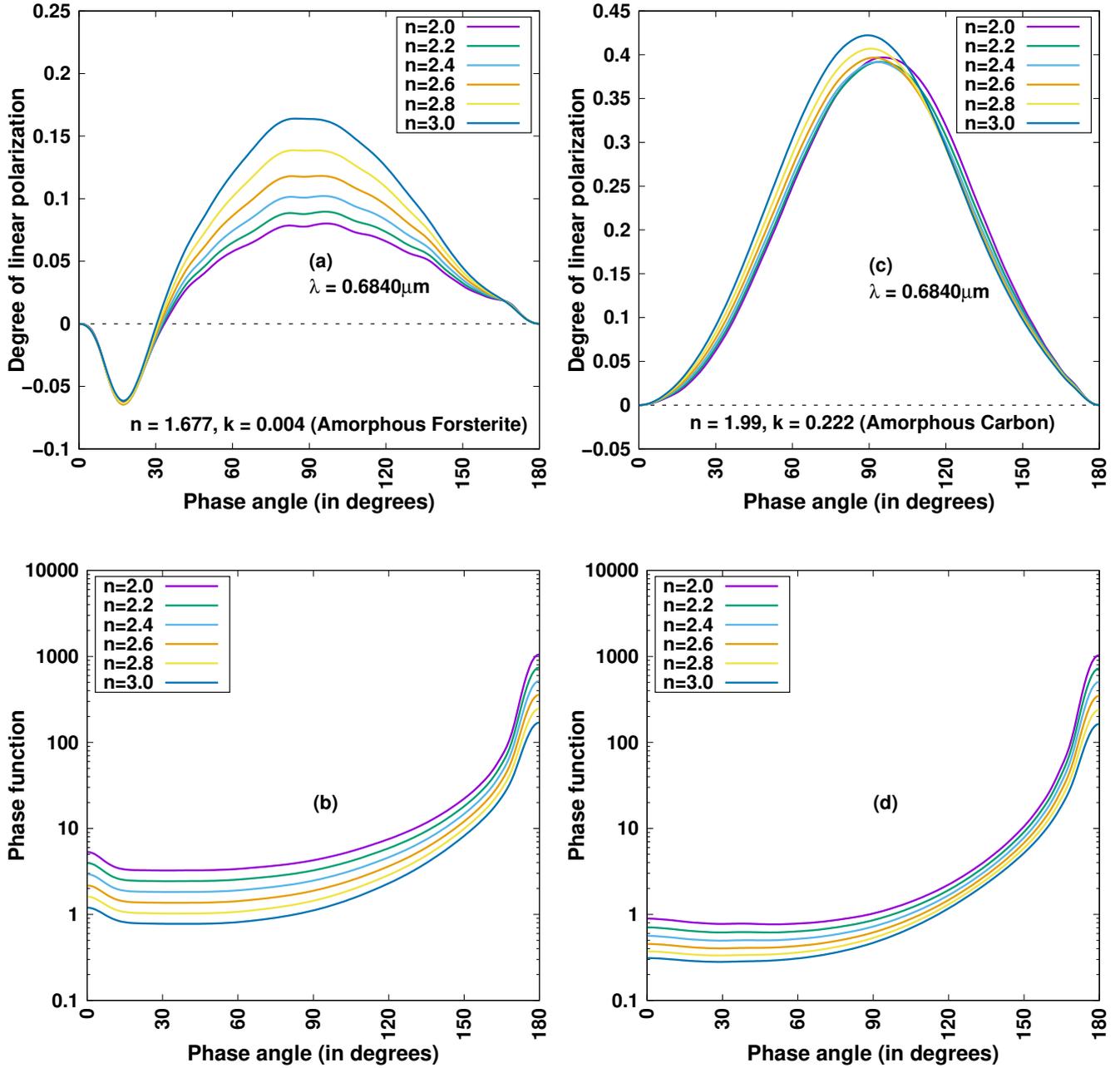}
    \caption{Variation of the degree of linear polarization and phase function for Fluffy Solids (FS) with increasing power-law index for amorphous forsterite (panels a \& b) and amorphous carbon (panels c \& d) respectively at $\lambda$ = 0.6840$\mu$m .}
    \label{fig:fs_pol_phase}
\end{figure*}

In this section we present the results of our light scattering simulations for the Fluffy Solids (FS) group of particles for silicate minerals and carbonaceous materials respectively.
Figure \ref{fig:fs_pol_phase} shows the variation in polarization and phase function with changing power-law size distribution index $n$ for silicate minerals and carbonaceous materials respectively at $\lambda$ = 0.6840 $\mu$m. Contrary to the polarimetric response obtained for the solids, the FS group of particles exhibit relatively narrow variation with $n$ both in the case of amorphous silicate \& amorphous carbon. Figure \ref{fig:fs_pol_phase}(a) also shows a steeper negative polarization with negligible dependence on $n$.
\subsection{Mixing of Hierarchical Aggregates \& Solids}
\begin{figure*}
	\includegraphics[width=\textwidth]{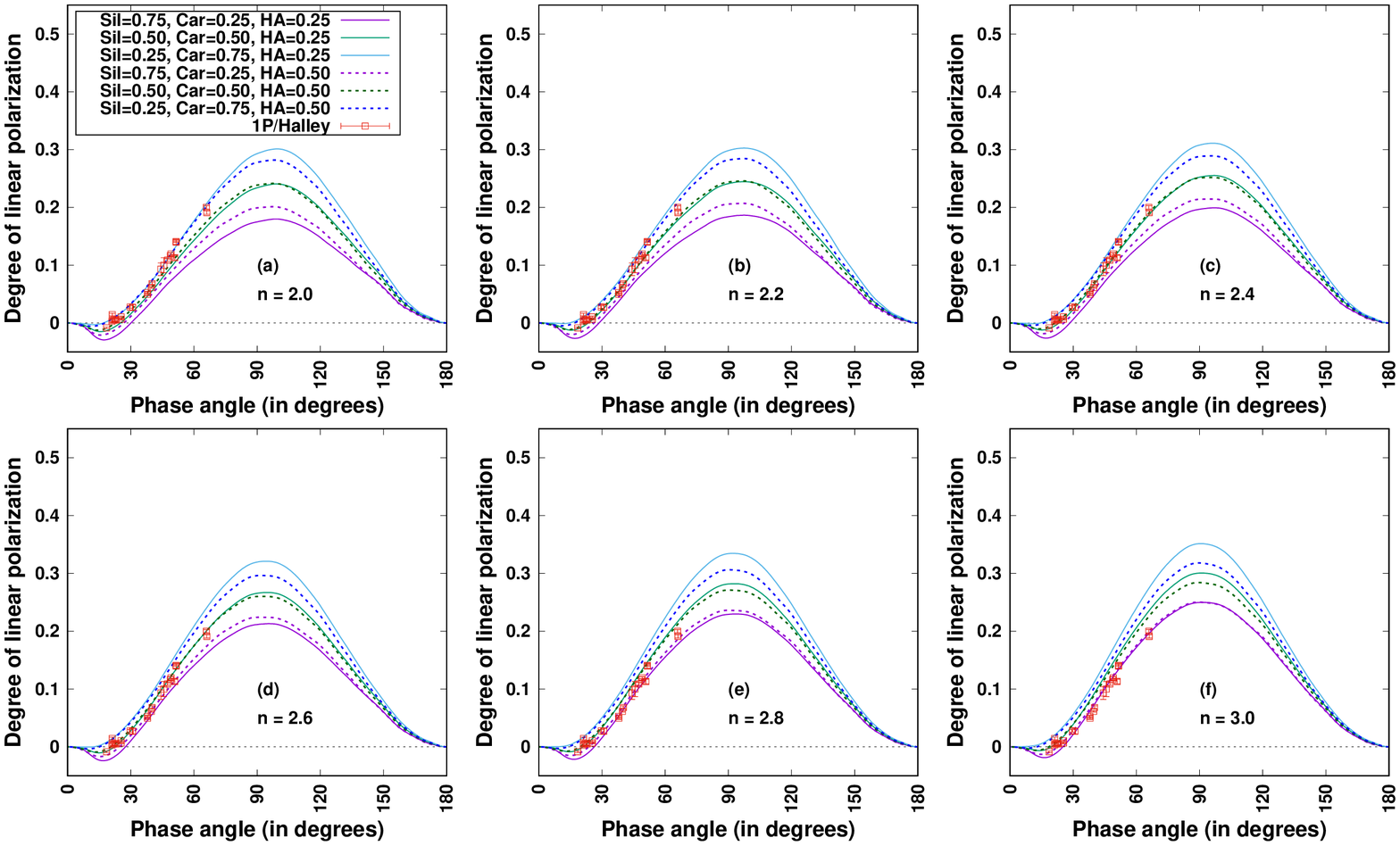}
    \caption{Variation of the degree of linear polarization with different mixing percentage of silicate and carbon for increasing power-law index $n$ = 2.0 (a), 2.2 (b), 2.4 (c), 2.6 (d), 2.8 (e) \& 3.0 (f) respectively at $\lambda$ = 0.6840$\mu$m under the effect of 25\% (solid lines) \& 50\% (dotted lines) added hierarchical aggregates (HA).}
    \label{fig:ha_ad_mix_n}
\end{figure*}

The Rosetta/MIDAS findings indicate the presence of both aggregates and solids, hence, in order to create a suitable comet dust model, it is necessary to include the contribution from hierarchical aggregates as well as that from the solid group of particles in the form of mixing percentages.
Figure \ref{fig:ha_ad_mix_n} depicts the variation in the degree of linear polarization for added 25\% and 50\% HA respectively with (25\% - 75\%) of solid silicates and carbonaceous materials for varying $n$.
One can easily notice from the figure, that for different mixing percentage of HA, there is significant variation in the polarimetric response. This particular figure exhibits a multi-dimensional approach as it depicts variation of the degree of linear polarization for changing $n$, silicate to carbon ratio and mixing percentage of HA. Following the same technique, we studied for mixing combination of HA + FS \& HA + FS + Solids at the narrow continuum wavebands.

\section{Comparison of model results with observations}
In this section we compare our model calculations with the observations {\citep{bastien1986, sen1991, chernova1993}} of several comets in all the narrow band continuum filters from the International Halley Watch (IHW)  program. 
The polarisation data for comets Halley, Hyakutake and Hale-Bopp were extracted from the database of cometary polarisation compiled by  \citet{kiselev2005}.  The observations of comet 67P/Churyumov-Gerasimenko are from \citet{myers1984, rosenbush2017}.

\subsection{Comet 67P/Churyumov-Gerasimenko}
\begin{figure}
	\includegraphics[width=8cm]{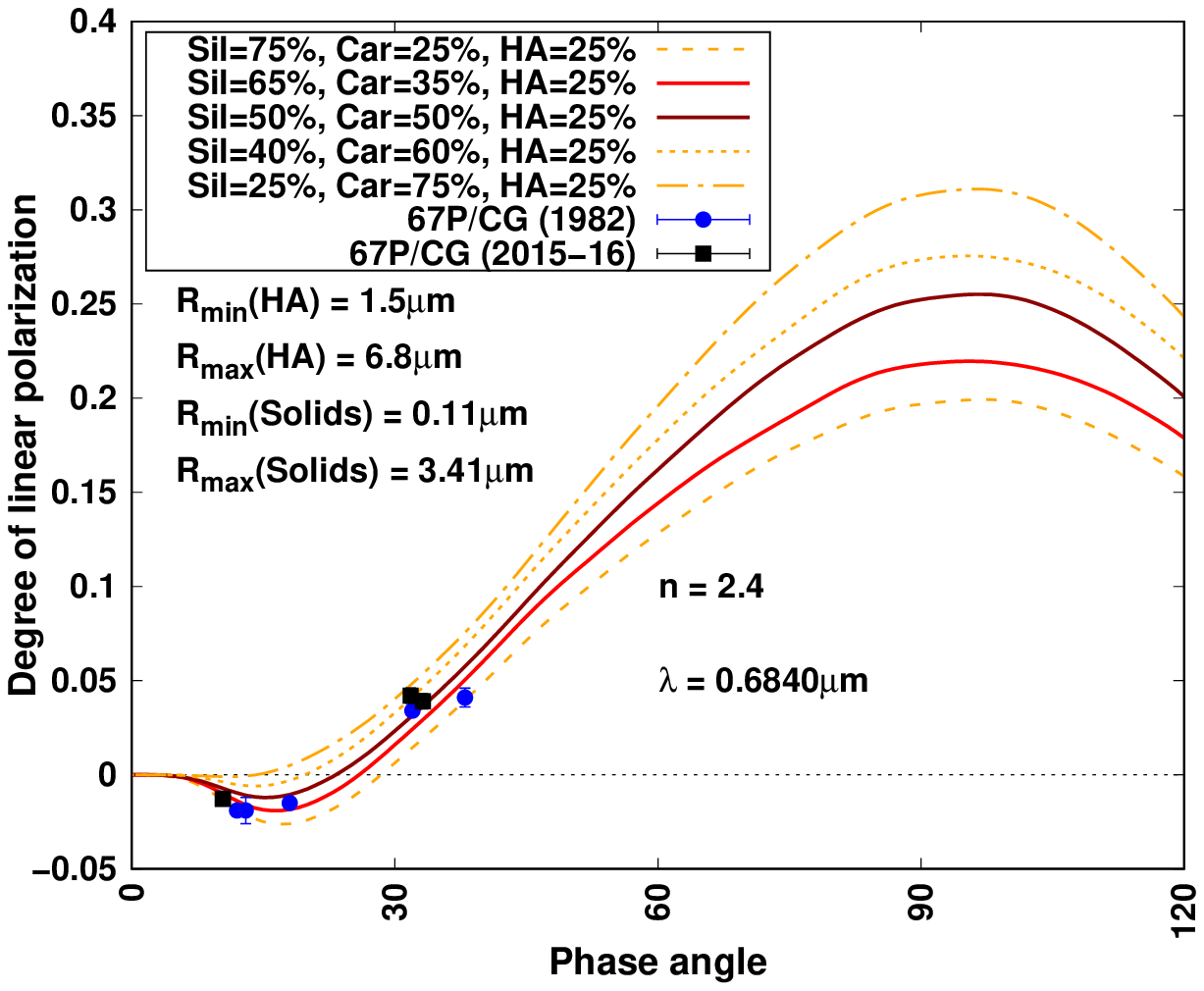}
    \caption{Best fit results for the comet 67P/Churyumov-Gerasimenko with the observations of 1982 \citep{myers1984} (blue circles) \& those from 2015-16 (black squares) \citep{rosenbush2017}}
    \label{fig:cg_fit}
\end{figure}

Figure \ref{fig:cg_fit} compares our model data with the observed polarisation from the 1982 \& 2015-16 apparitions for  comet 67P/Churyumov-Gerasimenko \citep{myers1984,rosenbush2017} in the red broadband filter. It is clear from the figure that the observed data seems to be fitted well by a mix of 50\% silicate minerals and 50\% carbonaceous material.

\subsection{Comet 1P/Halley}
\begin{figure*}
	\includegraphics[width=\textwidth]{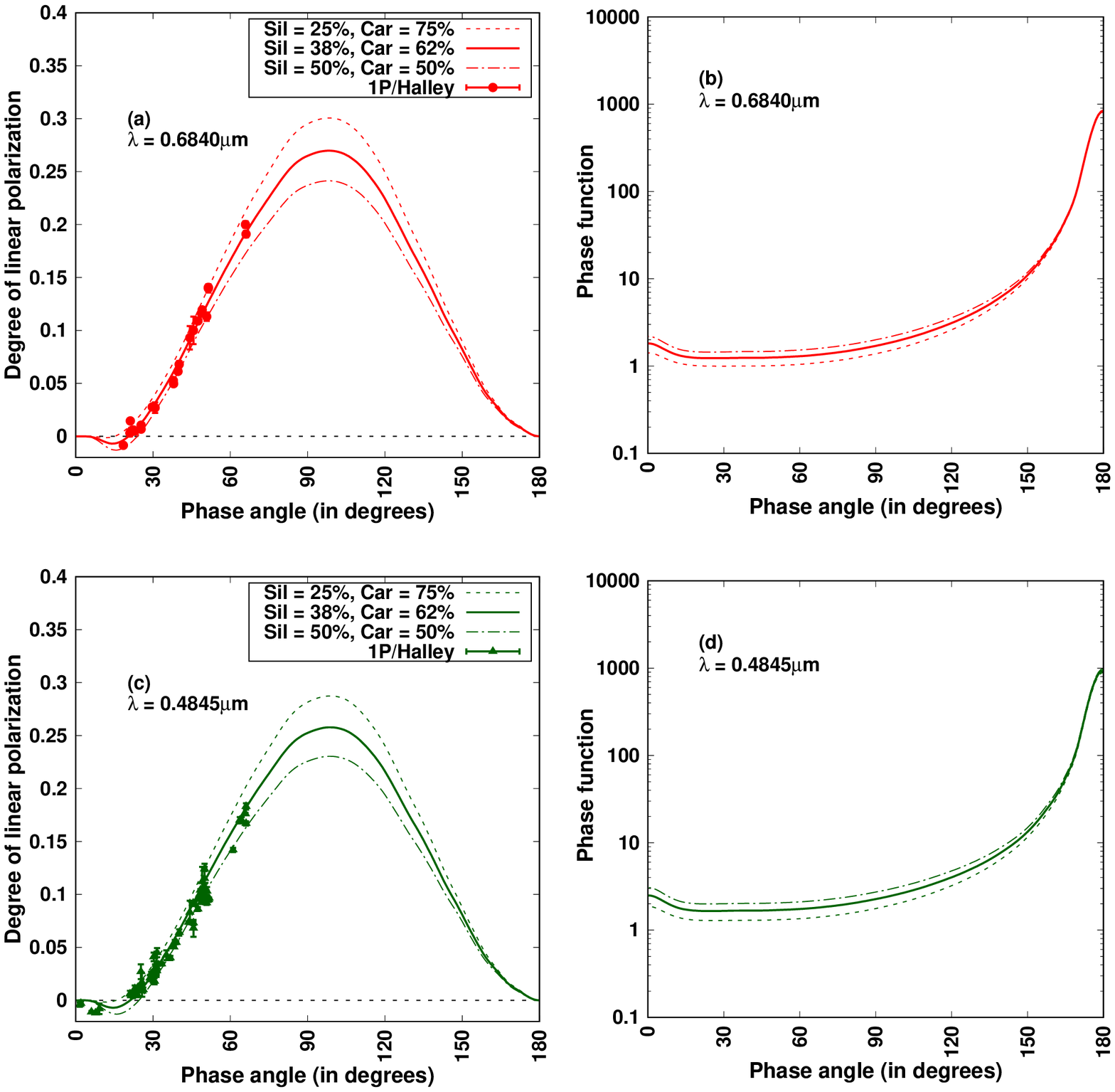}
    \caption{Best fit polarization results of the model for the comet 1P/Halley at $\lambda$ = 0.6840$\mu$m (a)\&(b) \& 0.4845$\mu$m (c)\&(d) respectively.}
    \label{fig:fit_halley}
\end{figure*}

\begin{figure*}
	\vspace{-80mm}
	\includegraphics[width=\textwidth]{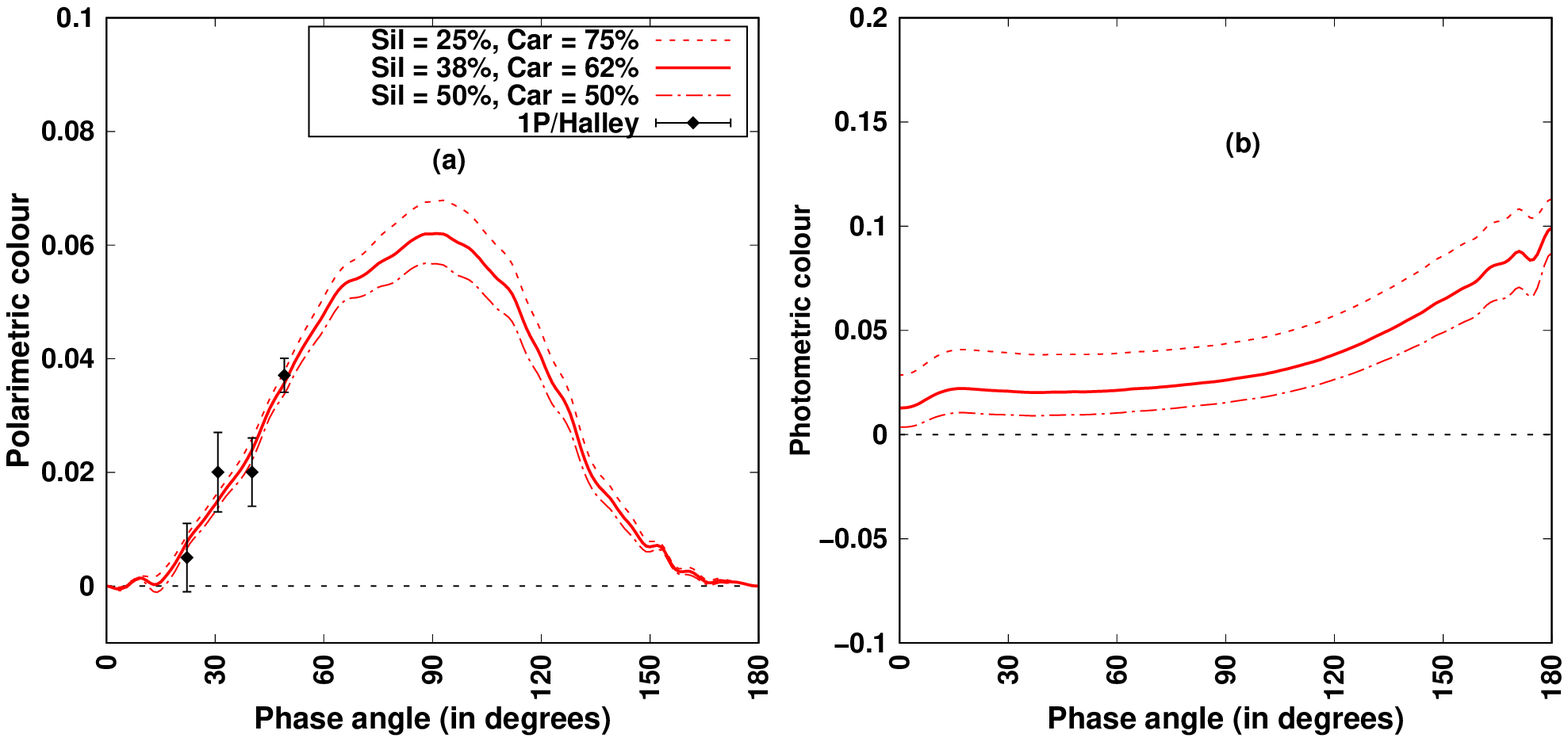}
    \caption{Polarimetric and photometric colour from the best fit model of comet 1P/Halley.}
    \label{fig:halley_col}
\end{figure*}

Figure  \ref{fig:fit_halley} shows the best fit curves obtained from the model for comet 1P/Halley at 0.6840$\mu$m \& 0.4845$\mu$m for different abundances of the silicates and carbonaceous material. The best fit results indicate the presence of 38\% of silicates and 62\% of carbonaceous material.   Some of the small number of outliers are better fitted within 25\% to 50\% of silicates.  We find that power-law index $n$ = 2.1 with the mixing combination of 25\% of HA and 75\% of Solids fits the observations acceptably. 
Thus, the model indicates the presence of relatively lower percentage of aggregates, moderate percentage of silicates and relatively large size particles in comet 1P/Halley which is in agreement with the literature \citep{kolkolova2010}. 
A comet dust model should explain photometric colour apart from the polarization properties.  Hence, we depict the color profiles, obtained from the model for comet 1P/Halley in Figure \ref{fig:halley_col}. Both the color profiles, polarimetric and photometric, have positive values and exhibit a similar trend as obtained by \cite{kolokolova2015} for their rough spheroid model. 

Figures  \ref{fig:fit_halebop_Hyakutake}(a), (b) \& (c) present the best fit results for the comet 1P/Halley at the three narrow band filters 0.6840$\mu$m, 0.4845$\mu$m \& 0.3650$\mu$m respectively. It is clear from the figures that the comet exhibits nearly shallow slope between red and green filters, whereas, the slope tends to be steeper as we approach the blue filter as shown in Figure \ref{fig:wavelength_dependence}(a). The observed dependence on wavelength is fitted well by the model for 25\% HA + 75\% Solids.  

\subsection{Comet C/1995 O1 (Hale-Bopp)}
The comet C/1995 O1 (Hale-Bopp) is a very unique long period comet having an orbital period of over 2000 years. The comet exhibits relatively high positive polarization as well as steep negative polarization which is usually not observed in most of the comets. Hence, it becomes difficult to model the polarimetric response of Hale-Bopp. Such unique polarimetric feature may arise due to presence of relatively small size sub-micron particles. We note that the combination of HA + Solids that are used to explain the observed polarimetric response from the short period comets, fail to explain those in case of the comet Hale-Bopp. As the comet falls in the category of long period comets, we considered the combination of HA + FS as discussed in section 2, in order to model the observed polarization at all the three narrow-band continuum filters as shown in the Figures  \ref{fig:fit_halebop_Hyakutake}(d),(e) \& (f) for $\lambda$ = 0.6840$\mu$m, 0.4845$\mu$m \& 0.3650$\mu$m respectively for 25\% HA + 75\% FS \& $n$=2.8. The best fit from the model, for all the three filters, are obtained for 35\% silicates \& 65\% carbon with few outliers ranging between 10\% - 90\% carbon. Although the observed data shows quite steep negative polarization, yet, the data points are accompanied by high percentage of error, hence the negative polarization in case of Hale-Bopp cannot be used to constrain the model parameters. 
However, our model indicates presence of high percentage of silicates at the negative polarization branch irrespective of the observational error.

Hale-Bopp exhibits relatively high variation in polarization as a function of wavelength with a steeper slope in the narrow band continuum filters compared to that exhibited by comet 1P/Halley as shown by the observations presented in \citet{ganesh1998}. We obtained similar wavelength dependence from our best fit model data for Hale-Bopp as shown in the Figure  \ref{fig:wavelength_dependence}(b) for various phase angles. Both the model and the observations show almost zero wavelength dependence at the negative polarization region. On the other hand as the phase angle increases, beyond the inversion angle, the wavelength dependence increases showing much steeper slope at 41.7$^{\circ}$ \& 46.6$^{\circ}$.   

\subsection{Comet C/1996 B2 (Hyakutake)}
\begin{figure*}
	\includegraphics[width=\textwidth]{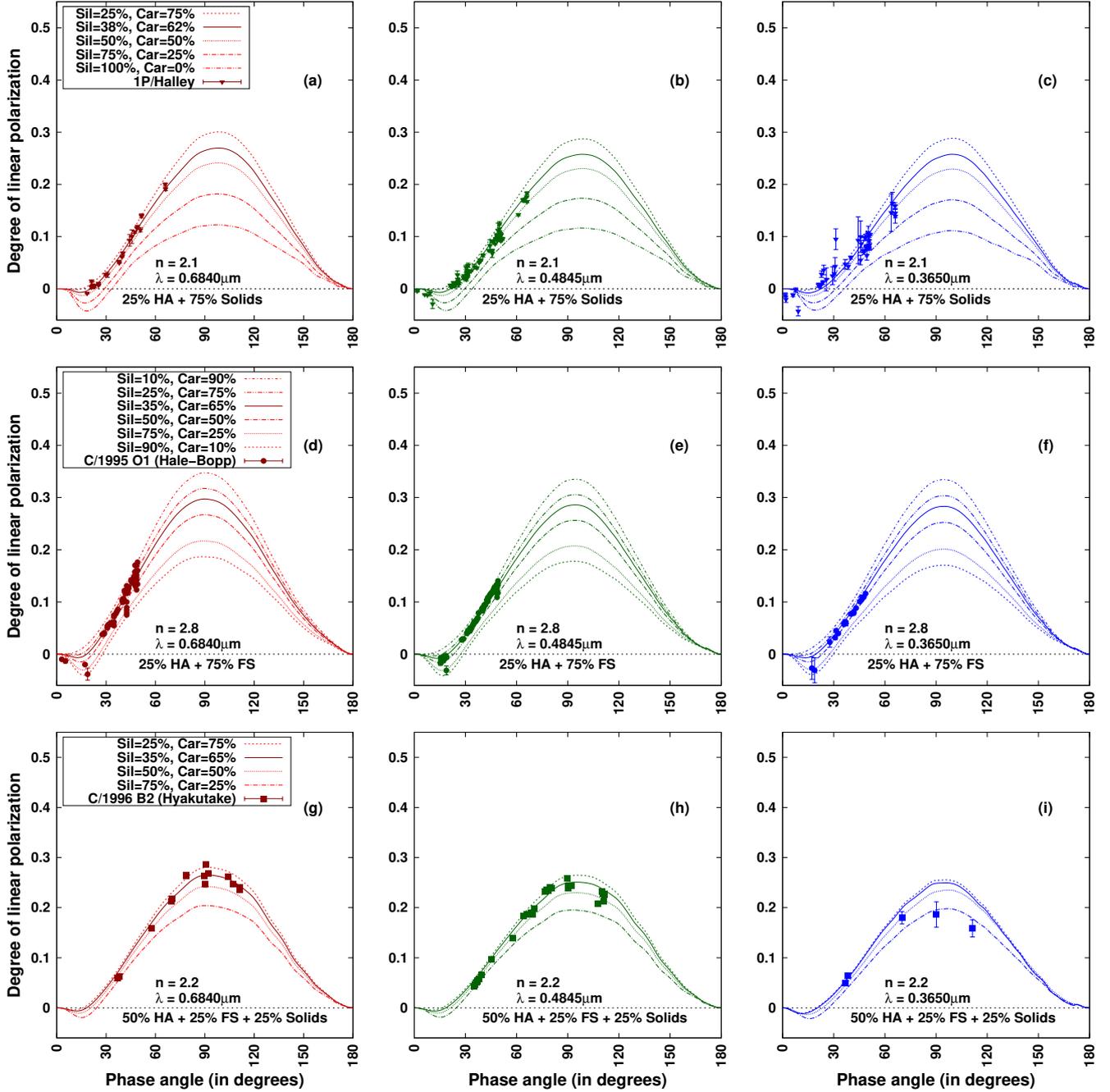}
    \caption{Best fit polarization reults from the model for the comets 1P/Halley [(a), (b) \& (c)], C/1995 O1 (Hale-Bopp) [(d), (e) \& (f)] and C/1996 B2 (Hyakutake) [(g), (h) \& (i)] at $\lambda$ = 0.6840$\mu$m, 0.4845$\mu$m \& 0.3650$\mu$m respectively }
    \label{fig:fit_halebop_Hyakutake}
\end{figure*}

Figure  \ref{fig:fit_halebop_Hyakutake}(g), (h) \& (i) depicts the best fit model data for the observed polarimetric response obtained for the comet C/1996 B2 (Hyakutake) at the three narrow band continuum filters $\lambda$ = 0.6840$\mu$m, 0.4845$\mu$m \& 0.3650$\mu$m respectively for 50\% HA + 25\% FS + 25\% Solids having $n$ = 2.2. We obtained the best fit results for 35\% Silicates \& 65\% Carbon at the red and green filters covering the entire phase range till 111$^{\circ}$. At the blue filter the number of observations are much less compared to those in the red \& green filters. Also, two of the observation points in the blue filter at phase angles 90.22$^{\circ}$ \& 111.36$^{\circ}$ exhibit higher percentage of error.  We also note that comet Hyakutake was quite active during this period and hence the coma dust properties may have varied over this period.  Hence, due to large error and less number of observation points accompanied by the uncertainty due to the activity, modelling Hyakutake at blue filter remains inconclusive. Despite the observational discrepancies in the blue filter, the wavelength dependence obtained from the model and the observations exhibit quite similar behaviour for all the three filters at phase angles 36.9$^{\circ}$, 38.1$^{\circ}$ and 70.22$^{\circ}$. At the phase angles 90.22$^{\circ}$ \& 111.36$^{\circ}$ the model and the observations are quite close for the red and green filters.

\subsection{Summary of the results}
Here, we summarise the best fit  results obtained from the model for the comets discussed above. 
The results are presented in Table - \ref{label:summary} for the four comets considered in this work.  The composition, dust morphology and the power law index for the PP size-distribution are listed in the table along with the orbital periods.  

We see that a mixture of hierarchical aggregates and solids is sufficient to explain the behaviour of polarization in the case of the short period comets, however, for the long period comets one needs to invoke solids with different porosity - i.e. the fluffy solids with varying percentages.   In each case, the power law index for the size distribution is different showing that the grain size distribution might be a specific characteristic of the comet.

\begin{table*}
	\caption{Summary of the best fit results obtained from the model for different comets along with their orbital periods}
	\begin{center}
\resizebox{15cm}{!}{
		\begin{tabular}{|c|c|c|c|c|}
			\hline
			\textbf{Comet} & \textbf{Composition} & \textbf{Morphology} & \textbf{n} & Orbital period\\
			\hline
\textbf{67P/Churyumov-Gerasimenko}	&	Sil = 50\%, Car = 50\%	&	25\% HA + 75\% Solids & 2.4 &	6.2 years \\
\textbf{1P/Halley}	&	Sil = 38\%, Car = 62\%	& 25\% HA + 75\% Solids & 2.1 & 75 years \\
\textbf{C/1995 O1 (Hale-Bopp)}	&	Sil = 35\%, Car = 65\%	& 25\% HA + 75\% FS & 2.8 & 2300 years \\
\textbf{C/1996 B2 (Hyakutake)}	&	Sil = 35\%, Car = 65\%	& 50\% HA + 25\% FS + 25\% Solids & 2.2	& 70,000 years \\
			\hline
		\end{tabular}}
	\end{center}
	\label{label:summary}
\end{table*}
\begin{figure*}
	\includegraphics[width=\textwidth]{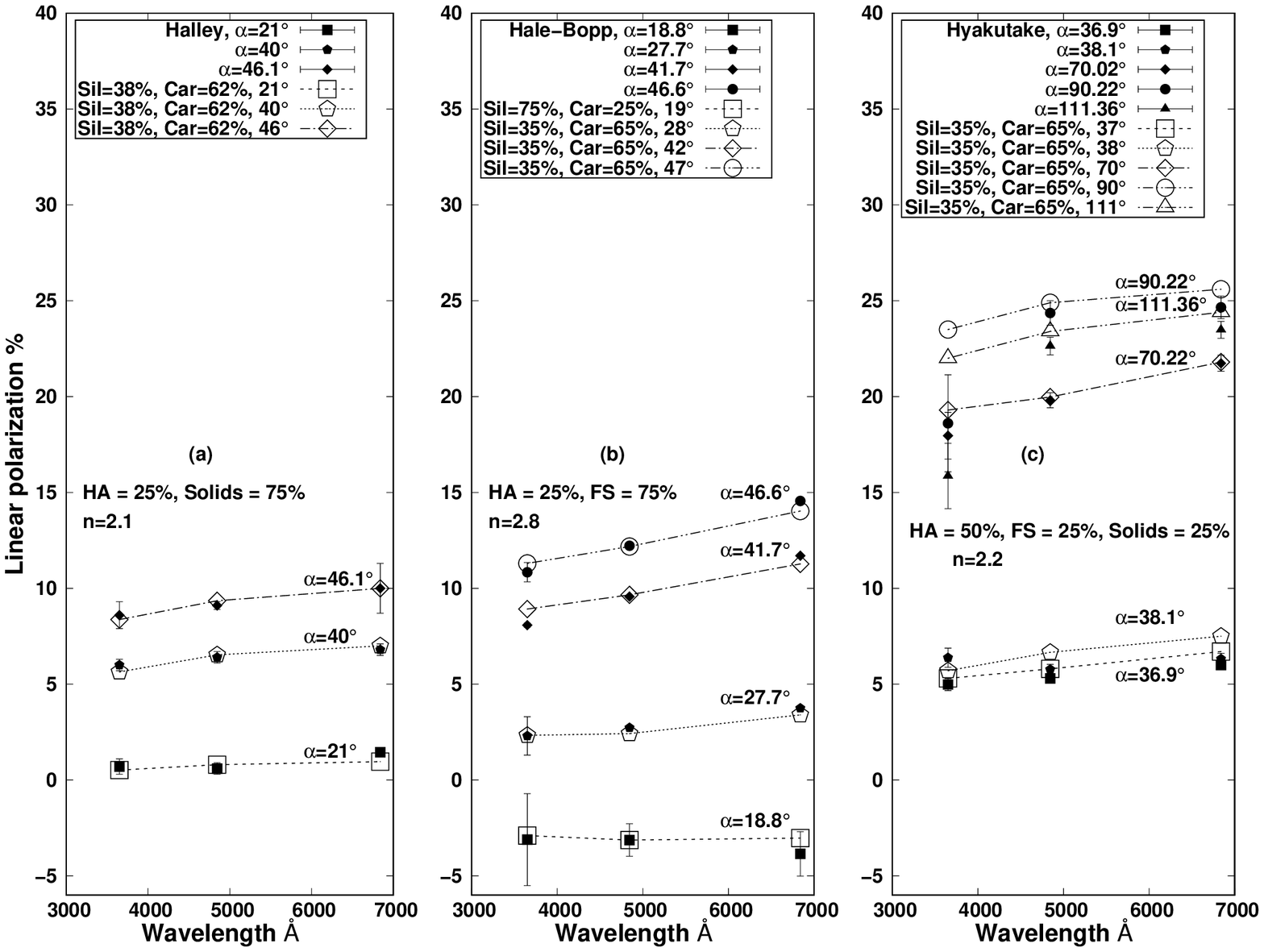}
    \caption{The best fit wavelength dependence of polarization for comet 1P/Halley (a), C/1995 O1 (Hale-Bopp) (b) \& C/1996 B2 (Hyakutake) (c) from the model data (hollow shapes) with the observed data (filled shapes) }
    \label{fig:wavelength_dependence}
\end{figure*}
\section{Discussion}
Under the framework of dust morphology revealed by the $Rosetta$/MIDAS instruments, we introduce a generalised cometary dust model that consider mixture of different dust morphologies as well as inhomogeneous mixture of silicate minerals and carbonaceous materials. The flexibility of the model allows us to control a large number of parameters like size, shape, porosity, composition and their mixing percentage as well as mixture of different morphologies like HA, FS and Solids and hence the same model is able to provide best fit results for different comets having different dynamical age. Also, the model results indicate presence of significant amount of HA \& FS in the case of long period comets, while the short period comets seem to fit well with relatively higher amount of Solids. Thus, it is clear from the model that long period comets are rich in high amount of loose particles (HA or FS or both) due to lesser magnitude of weathering. On the contrary, the short period comets have a larger amount of the solid (low porosity) particles - a signature that these group of comets have experienced higher magnitude of weathering. The recent modelling study of observed polarimetric response from protoplanetary disks in infrared, suggests presence of fluffy or loose particles \citep{tazaki2019} and hence we may consider, that the long period comets retain large amount of loose particles from the protoplanetary phase justifying our choice of dust morphology. The model polarimetric and photometric color (Figure  \ref{fig:halley_col}) profiles indicate positive trend as obtained by previous studies \citep{kolokolova2015}. Moreover, the study of wavelength dependence of polarization for the three comets (Figures  \ref{fig:fit_halebop_Hyakutake} \& \ref{fig:wavelength_dependence}) shows a good match between the observations and the model results mainly in the positive branch of polarization. But in Figures \ref{fig:fit_halebop_Hyakutake}[b, c \& d], the model data do not agree with the observations in the negative branch of polarization. The discrepancy in the fitting remains an open problem to be considered in a future work with specific attention to the negative branch of polarisation with additional observational data.    
Thus, the study of polarimetric color profiles and the wavelength dependence of polarization cross verifies the applicability of the model from multiple analytical viewpoints.   

\section{Conclusions}
Our findings from this work are briefly summarised below:

\begin{enumerate}[label=\roman*]
    \item The dust model presented in this work, introduces Hierarchical Aggregates(HA) having $D_f$ < 2 and considers Solids as agglomerated debris with porosity < 10\% having particle volume fraction of 23.6\%. The model also introduces a mixed morphology having Solid core and HA as filaments termed as Fluffy Solids (FS) having moderate porosity.  
    
    \item The variation of polydispersity reveals that with increasing $\sigma_p$ the slope of the polarization curve reduces from steeper to much shallower within the phase angles  20$^{\circ}$ to 90$^{\circ}$.  It attains the signature bell like shape which is found from various ground based polarimetric observations of comets. However, $D_2$ develops a peak with increasing $\sigma_p$ around the phase angle 30$^{\circ}$. Thus, an increase in $\sigma_p$ changes the anisotropy which in turn reduces the polarimetric slope keeping the phase function ($S_{11}$) unchanged.    
    
    \item The mixing of two different types of morphology allows the model to  introduce the contribution of both loose particles as well as solid or low porous particles. 
     
    \item For the comet 67P/Churyumov-Gerasimenko The model data fits with the observed polarization for a mix of 50\% silicate minerals and 50\% carbonaceous material under the influence of 25\% HA + 75\% Solids having $n$ = 2.4. 
     
    \item  The model data fits well with most of the observed polarization values of the comet 1P/Halley for 38\% silicates and 62\% carbonaceous materials and a mixing combination of 25\% HA + 75\% Solids with $n$ = 2.1. 

    \item In the case of the comet C/1995 O1 (Hale-Bopp) the model fits well with a majority of the observed points within the phase range of 30$^{\circ}$ to 55$^{\circ}$ for 35\% of silicates and 65\% of carbonaceous materials having 25\% HA + 75\% FS for n = 2.8, while a small number of observed polarization values agree with a model having 10\% silicates for phase angles beyond 50$^{\circ}$, while those outliers in the low phase angles require larger amount (50\% to 90\%) of silicates.
    
    \item For the comet C/1996 B2 (Hyakutake) the best fit model data is obtained for  35\% silicates and 65\% of carbonaceous materials having 50\% HA + 25\% FS + 25\% Solids for $n$ = 2.2. 
    
    \item Finally, our dust model 
    is verified with polarimetric color profile which is positive and exhibits similar trend as obtained by previous studies. Also, the model recreates the wavelength dependence of polarization.

\end{enumerate}

\section*{Acknowledgements}
We thank the referee, Dr. Dimitry Petrov, for the useful comments and suggestions which have improved the manuscript. We acknowledge Profs Daniel Mackowski and Michael Mishchenko, who made their Multi-sphere T-matrix code publicly available.  We also thank Prof. Asoke K. Sen \& Dr. Himadri Sekhar Das for their insights on cometary dust. We thank our colleagues at PRL for their comments.  We also thank Prof. Evgenij Zubko for useful discussion on comet dust modelling during the recent online meeting "International conference on Dust in Astrophysics (ICDA-2020)" held in the Dept. of Physics, Assam University, Silchar, India, during August 2020.  

Work at Physical Research Laboratory, Ahmedabad, is funded by the  Department of Space, Govt. of India. The authors acknowledge the use of the super-computing facility Vikram-HPC at PRL, Ahmedabad, where all the parallel computations related to this work were executed.  
\section*{Data Availability}

The observational data underlying this article are taken from various sources in the astronomical literature.  The simulated data from the theoretical models are available from the authors on reasonable request. 




\bibliographystyle{mnras}
\bibliography{manuscript_final} 


\bsp	
\label{lastpage}
\end{document}